\documentclass[conference]{IEEEtran}
\IEEEoverridecommandlockouts
\usepackage{cite}
\usepackage{amsmath,amssymb,amsfonts}
\usepackage{algorithmic}
\usepackage{graphicx}
\usepackage{textcomp}
\usepackage{xcolor}
\usepackage{cite}
\usepackage{float}
\usepackage{glossaries}
\def\BibTeX{{\rm B\kern-.05em{\sc i\kern-.025em b}\kern-.08em
		T\kern-.1667em\lower.7ex\hbox{E}\kern-.125emX}}

\newacronym{leo}{LEO}{Low-Earth Orbit}
\newacronym{pdf}{pdf}{Probability Density Function}
\newacronym{df}{DF}{Decode-and-Forward}
\newacronym{snr}{SNR}{Signal to Noise Ratio}   
\newacronym{nr}{NR}{New Radio}
\newacronym{ftr}{FTR}{Fluctuating Two-Ray}
\newacronym{ue}{UE}{User Equipment}
\newacronym{sr}{SR}{Shadowed-Rician}
\newacronym{csi}{CSI}{Channel Station Information}
\newacronym{los}{LOS}{Line-of-Sight}
\newacronym{ntn}{NTN}{Non-Terrestrial Networks}

\begin{document}
	\bstctlcite{IEEEexample:BSTcontrol}
	
	\title{Analysis of the Outage Probability of Ground-Based Relaying for Satellite Systems}
	
	\author{Hadi Hashemi, Beatriz Soret, M. Carmen Aguayo-Torres\thanks{Hadi Hashemi, Beatriz Soret and M. Carmen Aguayo-Torres (corresponding email: bsoret@ic.uma.es) are with the Telecommunication Research Institute (TELMA), Universidad de M\'{a}laga, Spain. }}

	\maketitle
	
	\begin{abstract}
		This paper investigates the theoretical basis for using ground relaying in multi-antenna satellites exposed to blocking situations. Inactive and unobstructed User Equipments (UEs) located on ground are the relaying nodes of UEs that are not in the field of view of the satellite. Exact closed-form relationships of the Signal-to-Noise Ratio (SNR) and the outage probability are obtained for the case where each user is connected to two transmitting antennas at the satellite. The channel between the satellite and ground is modeled as a shadowed-Rician (SR), whereas a Fluctuating Two-Ray (FTR) fading model is used for the mmWaves ground link between relay and UE, as well as cases with perfect and imperfect Channel State Information (CSI). The simulation results showed that if perfect CSI is not available in the pre-coding and only the signal phase is estimated, the performance loss is minimal and the system can reach its ideal performance by spending limited power. In any case, the closed-form for the ideal state are a good proxy to predict the performance under non-ideal conditions. 
	\end{abstract}

	\section{Introduction}
	Satellite communications are receiving a lot of attention in the last decade fueled by the boom of \gls{leo} communications and the quest for fully ubiquitous internet service. It is in the low orbits -- between $500$ km and $2000$ km -- where we have witnessed a rapid development of proprietary and legacy initiatives. 
	For the former, projects such as SpaceX’s Starlink, Amazon’s Kuiper and OneWeb
	are state-of-the-art examples of a new space market determined to bring internet
	to the most distant corners of the Earth. For the latter, the umbrella term of \gls{ntn} \cite{3GPPTR38.811} encompasses the 3GPP effort to include the whole air and space ecosystem in 5G \gls{nr} and the future 6G. In this vision, \gls{leo}  constellations with hundreds or even thousand of satellites are the enablers of extended cellular coverage,  serve as a global backbone, and offload  the cellular base stations in congested areas~\cite{3GPPTR22.822, TR38.913, 3GPPTR38.811, 3GPPTR38.821}. 
	
	However, the promise of ubiquitous service is challenged by the need of \gls{los} in satellite communications. As terrestrial \gls{ue}s move in a certain area, shadowing and multipath conditions might change abruptly causing large-scale variations in the received signal. This is critical in urban areas, with buildings and other obstacles that prevent line of sight between communicating devices, but also observable in rural low-scattered areas and depending on the elevation angle and the satellite move. One way to address Non-Line-of-Sight (NLOS) scenarios is to complement the satellite network with terrestrial relaying to ensure that terrestrial obstacles do not compromise the connectivity. The idea is that an unobstructed ground relay receives and retransmits signals between the satellite(s) and the obstructed ground users. At the same time, the trend towards higher frequencies on ground increases also the probability of interruption of the direct communication between source and destination. Therefore, relay networks play a key role in future terrestrial and non-terrestrial networks \cite{9328305} \cite{7109864}. 
	
Ground relaying has been widely addressed in the literature (see, e.g., \cite{laneman2004, conne2010}). Satellite relaying networks is in general a more recent topic, but several references have looked at different scenarios and challenges \cite{9789274,9789278,9776485,9774023,9744456}. Authors in \cite{9789274} consider a satellite as source and unmanned aerial vehicles (UAV) as relay to sent its data to user on earth. All nodes are equipped with a single antenna devices, and the outage probability of received signal to destination through decode-and-forward (DF) relay was studied. In \cite{9789278}, using UAVs as relay to control network when terrestrial relay infrastructure is unavailable, an approximate and asymptotic outage probabilities of the considered system in present interference are studied. The network performance of the satellite full-duplex relay with relay selectability is reviewed in \cite{9776485}. Authors in \cite{9774023} studied the physical-layer security of a hybrid satellite-terrestrial network where multiple terrestrial DF relays and users in the presence of a terrestrial eavesdropper. In \cite{9744456}, a two-user multi-relay non-orthogonal multiple access-based hybrid satellite-terrestrial relay network is proposed for minimizing the whole system outage probability and providing full diversity order for users.
 
Another key aspect in the analysis of terrestrial and non-terrestrial systems is the choice of a suitable and close to reality channel model. There are various statistical models for satellite channels and millimeter wave communication. Among them, we have chosen the shadowed-Rician (SR) fading model for the communication between the satellite and the user on the ground \cite{8747458} and the Fluctuating Two-Ray (FTR) fading model for the communication between the users on the ground \cite{7917287}. These models agree with the experimental results and thus bring the results of calculations in this article closer to reality \cite{1198102,7503970}. In recent years, the FTR model has been applied in the study of a relay network 
in the mode of amplify-and-forward as well as decoding-and-forward \cite{HASHEMI2020100991,9201014}. 
	
	In this paper, we study the theoretical basis for using ground relaying in multi-antenna satellites exposed to blocking situations. The decode-and-forward mode is assumed. After stating the problem model, the statistical behavior of the SNR and the probability of outage have been analyzed in two different cases for the amount of channel state information in the transmitter. 
	
	The rest of this article is organized as follows. In Section \ref{s2}, we present the system model. Section \ref{s3} presents our analyses to obtain the SNR distribution at the relay node when perfect and imperfect channel state information is present. Then in Sections \ref{s4}, we obtain the outage probability and, finally, Section \ref{s5} presents our numerical
	results followed by conclusions drawn in Section \ref{s6}.
	
	\section{System Model}\label{s2}
\begin{figure}
		\centering	\includegraphics[width=1.05\linewidth]{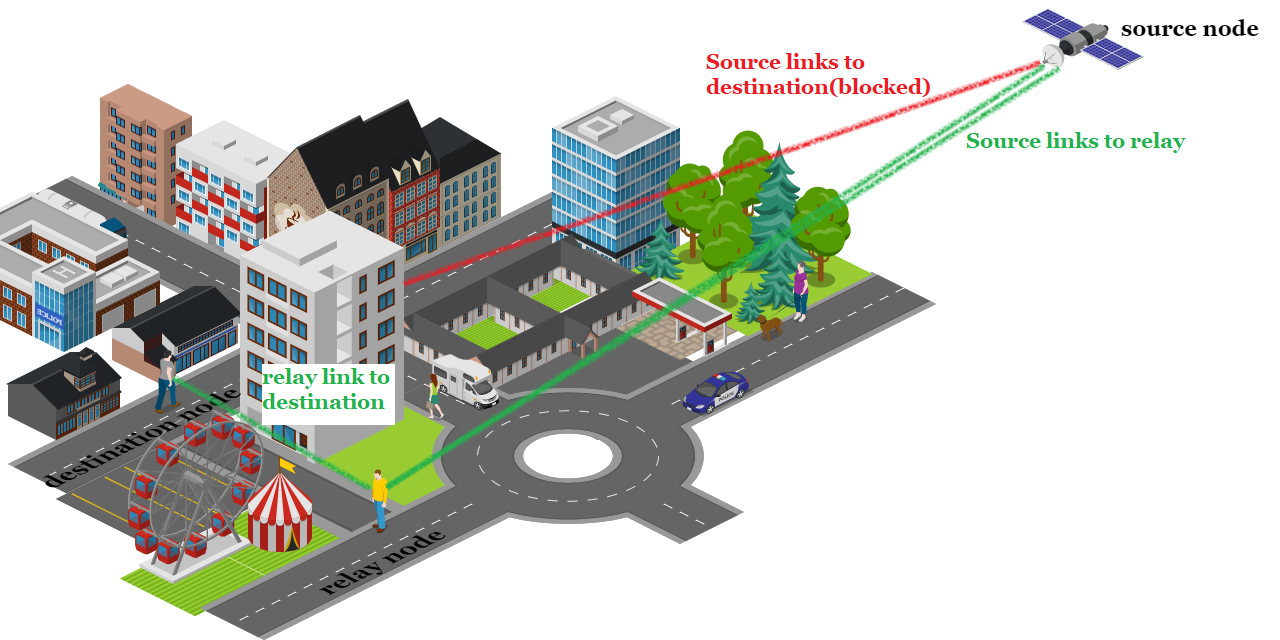}
		\caption{The system model of satellite communication through DF users.}
		\label{fig0}
	\end{figure}

	Consider the situation in Figure 
 ~\ref{fig0} where a LEO satellite with $N_s$ antennas provides service to single-antenna users located at the Earth surface within its coverage area. Assuming that the satellite knows the location of all users, the direction of the beam is set towards the intended user. As shown in the Figure, the end user receives a poor signal and cannot decode the message due to the blocking of its line of sight to the satellite. Another unobstructed node (which can be an inactive user or a ground station) acts as a relay to complete this communication. The received signal at the relay node can be modeled as follows
	\begin{align}
		y_{sr} = \sqrt{P_s G} \boldsymbol{h}^T_{sr} \boldsymbol{w} s + n,
	\end{align}
	Where $P_s$ is the power transmitted by each of the antennas, which can be optimally adjusted between the $N_s$ antennas, $G$ is the gain of the transmitter antennas in the direction of the desired user, $\boldsymbol{h}_{sr}$ is the channel vector between the source and the relay node,  $\boldsymbol{w}$ is a precoding vector that can be designed and specified based on the amount of existing \gls{csi}, $s$ is the desired signal, and $n$ is the additive zero-mean white Gaussian noise (AWGN) noise with power $N_0$. 
 
 Without losing the generality of the problem, we analyze the case of two antennas and assume that the power is equally divided between the two antennas, $P_s = \frac{P_t}{2}$. 
 We assume $\boldsymbol{h}_{sr}$ as the $2 \times 1$ shadowed-Rician (SR) channel vector between the source and the relay node. The complex base-band sample of a wireless fading channel for satellites can be expressed as
	\begin{align}
		h_{SR} = \zeta \sqrt{\Omega} e^{j\phi} + X + j Y, 
	\end{align}
	where $\zeta$ is a unit power Nakagami-m random variable for specular component with power $\Omega$ and a uniformly distributed random phase, $\phi$, such that $\phi \sim U[0, 2\pi)$. On the other hand, $X + j Y$ is a complex zero-mean Gaussian random variable with power $2\sigma^2$, such that $X,Y \sim \mathcal{CN}(0,\sigma^2)$. Based on these assumptions, the \gls{pdf} of $|h_{SR}|^2$ for a single link with SR fading is written~\cite{8747458}
	\begin{align}
		f_{|H_{SR}|^2}(x) = A e^{-\frac{x}{2\sigma^2}}  {}_1F_1(m,1,B x)
	\end{align}
	where $A = \frac{(2 \sigma^2 m)^m }{2\sigma^2(2 \sigma^2 m + \Omega)^m}$, $B = \frac{\Omega }{2\sigma^2(2 \sigma^2 m + \Omega)}$ , and ${}_1F_1 (·, ·, ·)$ denotes the confluent hyper-geometric function. This equation for integer $m$ can be represented as follows
	\begin{align}\label{pdf_sr_H2}
		f_{|H_{SR}|^2}(x) &= \sum_{k=0}^{m-1} \alpha_{SR}(k) x^k e^{-\beta_{SR} x}
	\end{align}
	where $(.)_k$ is the Pochhammer symbol, $\alpha_{SR}(k) =  \frac{m^m (-\Omega)^k (1-m)_k}{(2\sigma^2)^{k-m+1}  (2 \sigma^2 m + \Omega)^{k+m} (k!)^2}$ , and $\beta_{SR} = \frac{2 \sigma^2 m}{2\sigma^2(2 \sigma^2 m + \Omega)}$.
	
	Upon successful reception of the signal, the relay will decode it and send it to the destination. In other words, it works with the \gls{df} protocol. We can obtain the minimum required signal-to-noise ratio (SNR) level based on the data rate, $R$, and the bandwidth, $B$, and the decoding condition implies that the SNR in the relay is greater than $\eta_{th}=2^{R B}-1$. 
 
 The received signal from the relay at the destination node can be modeled as follows
	\begin{align}
		y_{rd} = \sqrt{P_r} h_{rd} s + n ,
	\end{align}
	where $P_r$ is the transmit power at the relay and $h_{rd}$ is the channel vector between the relay and the destination node. 

 The relay transmits this signal using a single antenna using ground millimeter waves. 
The channel between users is modeled according to the Fluctuating Two-Ray (FTR) fading model \cite{7917287}. We chose this model due to its high compatibility with the experimental results and its generality. Having adjustable parameters allows it to be accurately modeled for different conditions. In this model, in addition to the LOS signal, the largest environmental reflection is also considered. For this reason, the signal of this channel is given by
	\begin{align}
		h_{FTR} = \zeta V_1 e^{j\phi_1} +\zeta V_2 e^{j\phi_2} + X + j Y,
	\end{align}
	where the distribution of $\zeta$, $\phi_1$, $\phi_2$, $X$ and $Y$ are similar to the SR model. For this model, $K = \frac{V_1^2+V_2^2}{2\sigma^2}$ is the power ratio between specular and diffuse components, and $\Delta = \frac{2V_1V_2}{V_1^2+V_2^2}$ is the parameter expressing how similar to each other are the average received powers of the specular components. The pdf of the FTR model can be represented similarly to the SR model for integer $m$. The pdf of $|h_{FTR}|^2$ for a FTR fading with the use of \cite{7917287} and after some manipulation is as follows
	\begin{align}\label{pdf-h2-ftr}
		f_{|H_{FTR}|^2}(x) &= \sum_{i=1}^{M}\sum_{j=1}^{2}\sum_{b=0}^{m-1} \alpha_{FTR}(i,j,b) x^b e^{-\beta_{FTR}(i,j) x},
	\end{align}
	where $M = \lceil K \Delta \rceil + 1$, $\delta_i = \frac{\Delta\cos((i-1)\pi)}{2M-1}$, $\beta_{FTR}(i,j) = \frac{(K+1)m}{K(1+(-1)^j\delta_i)+m}$, $I = \int_{0}^{2M-1} \prod_{k=1,k\neq i}^{2M} (u-k+1) du$, and $\alpha_{FTR}(i,j,b) = \frac{m^m (K+1)^{b+1} (-1)^i (K(1+(-1)^j\delta_i))^b \binom{m-1}{b} I}{(K(1+(-1)^j\delta_i)+m)^{m+b} (2M-1) \Gamma(2M-i+1) \Gamma(i) \Gamma(b+1)}$. By adjusting the parameters of this model, all other fading models can be reached \cite{7917287}. Therefore, the analysis of this model will be very valuable.
	\\
	\section{Probability Density Function of The SNR at the Relay Node} \label{s3}
	To obtain the outage probability, we first have to derive the pdf of the SNR at the relay node. 
 According to the level of CSI, the source node can choose the appropriate precoding vector so that the SNR at the relay node is maximized. 
 We consider two different CSI scenarios: perfect CSI and imperfect (partial) CSI. 
	
	\subsection{Perfect CSI}
 With perfect channel estimation, we consider the precoding vector for the satellite as follows
	\begin{align}
		\boldsymbol{w} = \frac{\boldsymbol{h}_{sr}^*}{\sqrt{||\boldsymbol{h}_{sr}||_2}}
	\end{align}
	Therefore, the received signal is
	\begin{align}
		y_{sr} &= \sqrt{P_s G} \boldsymbol{h}^T_{sr} \frac{\boldsymbol{h}^*_{sr}}{\sqrt{||\boldsymbol{h}_{sr}||_2}} s + n  
		= \sqrt{P_s G ||\boldsymbol{h}_{sr}||_2} s + n .
	\end{align}
	The SNR at relay node is $\Lambda_r = \frac{P_s G (|h_{sr,1}|^2+|h_{sr,2}|^2 )}{N_0} = \Lambda_{r,1}+\Lambda_{r,2}$, where $\Lambda_{r,i} = \frac{P_s G |h_{sr,i}|^2}{N_0} = \bar{\rho} |h_{sr,i}|^2$ is the received SNR from $i$th channel between the source and relay node. With the use of \eqref{pdf_sr_H2}, the pdf of the received SNR from each link can be written
	\begin{align}
		f_{\Lambda_{r,i}}(x) &=  \frac{1}{\bar{\rho}} f_{|H_{sr,i}|^2}(\frac{x}{\bar{\rho}}) = \sum_{k=0}^{m-1} \alpha'_{SR}(k) x^k e^{- \beta'_{SR} x},
	\end{align}
	where $\alpha'_{SR}(k) = \frac{\alpha_{SR}(k)}{\bar{\rho}^{k+1}}$ and $\beta'_{SR} = \frac{\beta_{SR}}{\bar{\rho}}$. Due to the small distance of the antennas to each other (as much as half a wavelength) and the fact that both links have similar conditions, it can be assumed that the parameters of these two links are similar but they are independent. Because the total SNR is the sum of two independent SNRs, we apply the distribution of the sum of the independent random variables. For this we do the convolution of their distribution functions together. With two active antennae, and according to \cite[Eq 3.191.1]{zwillinger2007table}, this yields
	\begin{align}\label{pdf-snr-relay-perfect}
		&f_{\Lambda_r}(x) = f_{\Lambda_{r,1}}*f_{\Lambda_{r,2}}(x) \\ 
		&= \sum_{k_1=0}^{m-1} \sum_{k_2=0}^{m-1} \alpha'_{SR}(k_1)\alpha'_{SR}(k_2) e^{-\beta'_{SR} x} \int_{0}^{x}  t^{k_1}(x-t)^{k_2}dt \nonumber\\ 
		&= \sum_{k_1=0}^{m-1} \sum_{k_2=0}^{m-1} \alpha'_{SR}(k_1)\alpha'_{SR}(k_2) B(k_1+1,k_2+1)  x^{k_1+k_2+1} e^{-\beta'_{SR} x} \nonumber
	\end{align}
	where $B(.,.)$ is the Beta function.
	
	\subsection{Imperfect CSI}
	It is more difficult to estimate the amplitude of the channel than its phase. In this subsection, we suppose that the satellite has only an estimation of the channel phases. Then, the precoding vector is as follows
	\begin{align}
		\boldsymbol{w} = \frac{1}{\sqrt{2}} [e^{-j\theta_1},e^{-j\theta_{2}}]^T.
	\end{align}
	With the use of this vector, the received signal is similar to the output of the equal gain combiner, i.e., 
	\begin{align}
		y_{sr} 	& = \sqrt{\frac{P_s G}{2}} \left( |h_{sr,1}|+|h_{sr,2}| \right)  s + n.
	\end{align}
	Let us define $h_{t} = |h_{sr,1}|+|h_{sr,2}|$, so the SNR at relay node is $\Lambda_r = \frac{\bar{\rho}}{2} h^2_{t}$. As expected, by reducing the level of information from the channel, the power level reached to the relay also decreases, because the relationship $ \frac{\left( |h_{sr,1}|+|h_{sr,2}| \right)^2}{2} < |h_{sr,1}|^2+|h_{sr,2}|^2$ is always established. 
 The pdf of the channel amplitude is $f_{|H|}(x) = 2 x f_{|H_{SR}|^2}(x^2)$. Unlike the perfect CSI case, we have to work with the convolution of two pdfs of the channel amplitudes as follows
	\begin{align}
		&f_{H_t}(x) = f_{H_{sr,1}}*f_{H_{sr,2}} (x) \nonumber\\ 
		&= \sum_{k_1=0}^{m-1} \sum_{k_2=0}^{m-1} 4 \alpha_{SR}(k_1) \alpha_{SR}(k_2)  \\
		&~~~~~~~\times \int_{0}^{x} t^{2k_1+1}(x-t)^{2k_2+1} e^{-\beta_{SR} (t^2 + (x-t)^2)} dt.\nonumber
	\end{align}
	By change of variables $u = t - \frac{x}{2}$ and binomial expansions of the resulting terms, we get
	\begin{align}
		&f_{H_t}(x) = \sum_{k_1=0}^{m-1} \sum_{k_2=0}^{m-1} \sum_{b_1=0}^{2k_1+1} \sum_{b_2=0}^{2k_2+1} 4 \alpha_{SR}(k_1) \alpha_{SR}(k_2) e^{-\beta_{SR} \frac{x^2}{2}} \nonumber \\
		& \binom{2k_1+1}{b_1}\binom{2k_2+1}{b_2} (\frac{x}{2})^{2k_1+2k_2+2-b_1-b_2} (-1)^{b_2} I,
	\end{align}
	where $I = \int_{-\frac{x}{2}}^{\frac{x}{2}} u^{b_1+b_2} e^{-2\beta_{SR} u^2} dt$. This integral for odd values of $b_1+b_2$ is zero. We can rewrite integral as $I = ((-1)^{b_1+b_2}+1) \int_{0}^{\frac{x}{2}} u^{b_1+b_2} e^{-2\beta_{SR} u^2} dt$. Because this relation is the definition of a lower incomplete gamma, $I = \frac{(-1)^{b_1+b_2}+1}{2 (\sqrt{2 \beta_{SR}})^{b_1+b_2+1}} \gamma(\frac{b_1+b_2+1}{2}, \frac{\beta_{SR} x^2}{2})$, and we can rewrite
	\begin{align} \label{eq:Ht}
		&f_{H_t}(x) = \sum_{k_1=0}^{m-1} \sum_{k_2=0}^{m-1} \sum_{b_1=0}^{2k_1+1} \sum_{b_2=0}^{2k_2+1} \alpha_{SR}(k_1) \alpha_{SR}(k_2)  \nonumber \\
		& \binom{2k_1+1}{b_1}\binom{2k_2+1}{b_2}  \frac{(-1)^{b_1}+(-1)^{b_2}}{2^{2k_1+2k_2+1-\frac{b_1+b_2-1}{2}}} \frac{1}{\beta^{\frac{b_1+b_2+1}{2}}_{SR}} \nonumber\\
		& x^{2k_1+2k_2+2-b_1-b_2} e^{-\beta_{SR} \frac{x^2}{2}} \gamma(\frac{b_1+b_2+1}{2}, \frac{\beta_{SR} x^2}{2}).
	\end{align}
	The pdf of the SNR is related to the pdf of the channel amplitude as follows
	\begin{align}
		f_{\Lambda_r}(x) = \frac{1}{\sqrt{2 \bar{\rho} x}} f_{H_t}\left( \sqrt{\frac{2 x}{\bar{\rho}}}\right) .
	\end{align}
	And using (\ref{eq:Ht}) we write
	\begin{align}\label{pdf-snr-relay-imperfect}
		&f_{\Lambda_r}(x) = \sum_{k_1=0}^{m-1} \sum_{k_2=0}^{m-1} \sum_{b_1=0}^{2k_1+1} \sum_{b_2=0}^{2k_2+1} \alpha'_{SR}(k_1) \alpha'_{SR}(k_2)  \nonumber \\
		& \binom{2k_1+1}{b_1}\binom{2k_2+1}{b_2}  \frac{(-1)^{b_1}+(-1)^{b_2}}{2^{k_1+k_2+1}} \frac{1}{\beta'^{\frac{b_1+b_2+1}{2}}_{SR}} \nonumber\\
		& x^{k_1+k_2-\frac{b_1+b_2-1}{2}} e^{-\beta'_{SR} x} \gamma(\frac{b_1+b_2+1}{2},\beta'_{SR} x).
	\end{align}
	
	\section{Outage Probability Analysis}\label{s4}

 Outage in the destination can occur under two conditions. First, the signal power in the satellite path is lower than the threshold and the relay cannot receive the correct signal. The second may occur when the relay does not deliver enough power to receive the message. We define this total outage probability in the network as follows
	\begin{align}
		P_{Out}(\eta_{th}) &= P_{Out-SR}(\eta_{th}) \nonumber \\
		&~~~~+ (1-P_{Out-SR}(\eta_{th})) P_{Out-FTR}(\eta_{th}),
	\end{align}

\noindent where $P_{Out-SR} = Pr(\Lambda_r \leq \eta_{th}) $ is the outage probability in the source-relay link, $P_{Out-FTR} = Pr(\Lambda_d \leq \eta_{th}) $ is the outage probability in the relay-destination link, and $\eta_{th}$ is a pre-defined threshold. The closed-forms of $P_{Out-SR}$ and $P_{Out-FTR}$ are obtained next.
 

	
 \subsection{Outage Possibility of the Satellite Link with Perfect CSI}
	When perfect CSI is available, we have a pdf of SNR at the relay node from \eqref{pdf-snr-relay-perfect}. The outage probability can be studied as follow
	\begin{align}
		&P_{Out-SR}(\eta_{th}) = \int_{0}^{\eta_{th}} f_{\Lambda_r}(x) dx \nonumber\\ 
		&= \sum_{k_1=0}^{m-1} \sum_{k_2=0}^{m-1} \alpha'_{SR}(k_1)\alpha'_{SR}(k_2) B(k_1+1,k_2+1)  \nonumber\\
		& ~~~~~~~~~~~~~~~~\times \int_{0}^{\eta_{th}} x^{k_1+k_2+1} e^{-\beta'_{SR} x} dx
	\end{align}
	According to \cite[Eq 8.350.1]{zwillinger2007table}  we have
	\begin{align}\label{outsr1}
		P_{Out-SR}(\eta_{th}) &=  \sum_{k_1=0}^{m-1} \sum_{k_2=0}^{m-1} \alpha'_{SR}(k_1)\alpha'_{SR}(k_2) B(k_1+1,k_2+1)  \nonumber\\
		& ~~~~~~~~~~~~~\times \frac{\gamma(k_1+k_2+2,\beta'_{SR}\eta_{th} )}{\beta'^{k_1+k_2+2}_{SR}} 
	\end{align}

	\subsection{Outage Possibility of the Satellite Link with Imperfect CSI}
	The distribution function of the SNR changes in this situation. On average, we will be a little far from the optimal power and it is expected that the probability of outage is higher. The outage probability is given by \eqref{pdf-snr-relay-imperfect}
	\begin{align}
		&P_{Out-SR}(\eta_{th}) = \int_{0}^{\eta_{th}} f_{\Lambda_r}(x) dx \nonumber\\ 
		& = \sum_{k_1=0}^{m-1} \sum_{k_2=0}^{m-1} \sum_{b_1=0}^{2k_1+1} \sum_{b_2=0}^{2k_2+1} \alpha'_{SR}(k_1) \alpha'_{SR}(k_2)  \nonumber \\
		& \binom{2k_1+1}{b_1}\binom{2k_2+1}{b_2}  \frac{(-1)^{b_1}+(-1)^{b_2}}{2^{k_1+k_2+1}} \frac{1}{\beta'^{\frac{b_1+b_2+1}{2}}_{SR}} \nonumber\\
		& \times \int_{0}^{\eta_{th}}x^{k_1+k_2-\frac{b_1+b_2-1}{2}} e^{-\beta'_{SR} x} \gamma(\frac{b_1+b_2+1}{2},\beta'_{SR} x)  dx
	\end{align}
	To solve this integral, we use the series form of the gamma function, $\gamma(a,x)=  \sum_{d = 0}^{\infty} \frac{\Gamma(a)}{\Gamma(a+d+1)} x^{a+d} e^{-x}$, and after some rewriting we obtain
	\begin{align}\label{outsr2}
		&P_{Out-SR}(\eta_{th}) = \int_{0}^{\eta_{th}} f_{\Lambda_r}(x) dx \nonumber\\ 
		& = \sum_{k_1=0}^{m-1} \sum_{k_2=0}^{m-1} \sum_{b_1=0}^{2k_1+1} \sum_{b_2=0}^{2k_2+1} \sum_{d = 0}^{\infty} \alpha'_{SR}(k_1) \alpha'_{SR}(k_2)  \nonumber \\
		& \binom{2k_1+1}{b_1}\binom{2k_2+1}{b_2}  \frac{(-1)^{b_1}+(-1)^{b_2}}{2^{2k_1+2k_2+d+3}} \frac{\Gamma(\frac{b_1+b_2+1}{2}) }{\Gamma(\frac{b_1+b_2+1}{2}+d+1)}  \nonumber\\
		& \times \frac{\gamma(k_1+k_2+d+2,2\beta'_{SR}\eta_{th} )}{\beta'^{k_1+k_2+2}_{SR}} 
	\end{align}
	
	\begin{figure}
		\centering
		\includegraphics[width=1.05\linewidth]{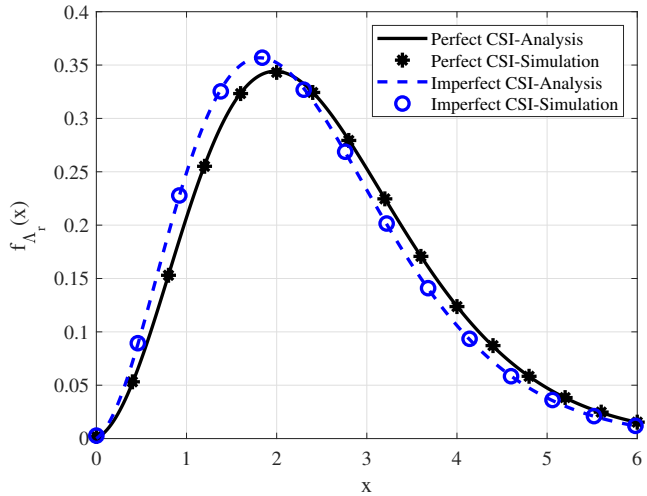}
		\caption{The probability density function of the SNR at the relay node when CSI is perfect or imperfect.}
		\label{fig1}
	\end{figure}
	
	\subsection{Outage Possibility of the Ground Link}
 For the millimiter link between the ground relay and the ground destination, the SNR distribution function, $f_{\Lambda_d}(x) = \frac{1}{\bar{\rho}} f_{|H_{FTR}|^2} \left(\frac{x}{\bar{\rho}}\right) $, is written using \eqref{pdf-h2-ftr}
	\begin{align}
		f_{\Lambda_d}(x) &= \sum_{i=1}^{M}\sum_{j=1}^{2}\sum_{b=0}^{m-1} \alpha'_{FTR}(i,j,b) x^b e^{-\beta'_{FTR}(i,j) x}
	\end{align}
	where $\alpha'_{FTR}(i,j,b) = \frac{\alpha_{FTR}(k)}{\bar{\rho}^{b+1}}$ and $\beta'_{FTR} = \frac{\beta_{FTR}}{\bar{\rho}}$. Then, the outage probability can be formulated as follows
	\begin{align}
		P_{Out-FTR}(\eta_{th}) = \sum_{i=1}^{M}\sum_{j=1}^{2}\sum_{b=0}^{m-1}  \frac{\alpha'_{FTR}(i,j,b)}{\beta'^{b+1}_{FTR}} \gamma(b+1,\beta'_{FTR}\eta_{th} )
	\end{align}

	\begin{figure}
		\centering
		\includegraphics[width=1.05\linewidth]{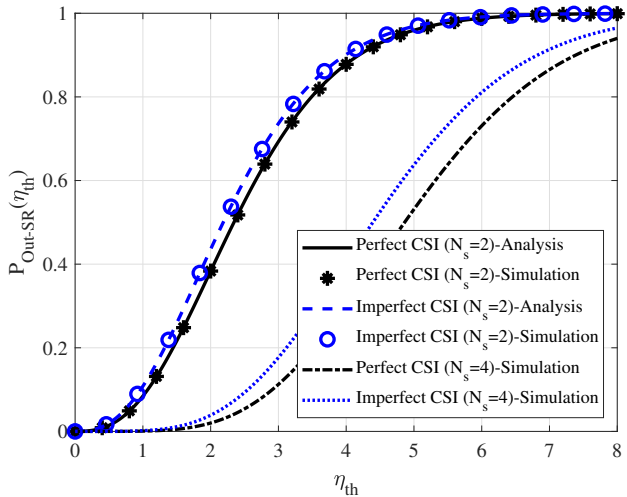}
		\caption{The outage probability versus threshold at the relay node when CSI is perfect or imperfect whit 2 or 4 active antenna at the source.}
		\label{fig2}
	\end{figure}
	\section{Simulation Results}\label{s5}
	
	In this section, we provide numerical results to validate the analysis and quantify the system performance. Monte Carlo simulations have been done with $10^6$ iterations using MATLAB software. 
	
	Figure \ref{fig1} shows the pdf of the SNR at the relay node with perfect and imperfect CSI. For each path to the relay node, the power budget is set to $P_s=1$, the gain of the transmitter antennas is one and the average SNR is $1$ dB. The channel gains are normalized by setting the expected value of their squares to 1 and $m=5$. 
 With perfect CSI the peak of the graph is tilted to the right, which means that more estimation of the channel increases the communication quality. The average received power for this simulation is 2.51 dB in perfect CSI and 2.35 dB in imperfect CSI. 
	
	In Figure \ref{fig2}, the outage probability versus the threshold at the relay node when the CSI is perfect or imperfect is depicted. Besides the analyzed case of 2 antennas, we show the simulated result with 4 active antennae at the source to give additional insight of the generalization of our analyses. 
 For the case with two antennae, the figure shows the accuracy of equations \eqref{outsr1} and \eqref{outsr2}. The series in relation \eqref{outsr2} is truncated to 30 terms. Having perfect CSI leads to a lower outage probability as compared to imperfect CSI, and this improvement is more visible as the number of antennae increases. As expected, increasing the number of antennae reduces the outage probability. 
	
\begin{figure}
		\centering
		\includegraphics[width=1.05\linewidth]{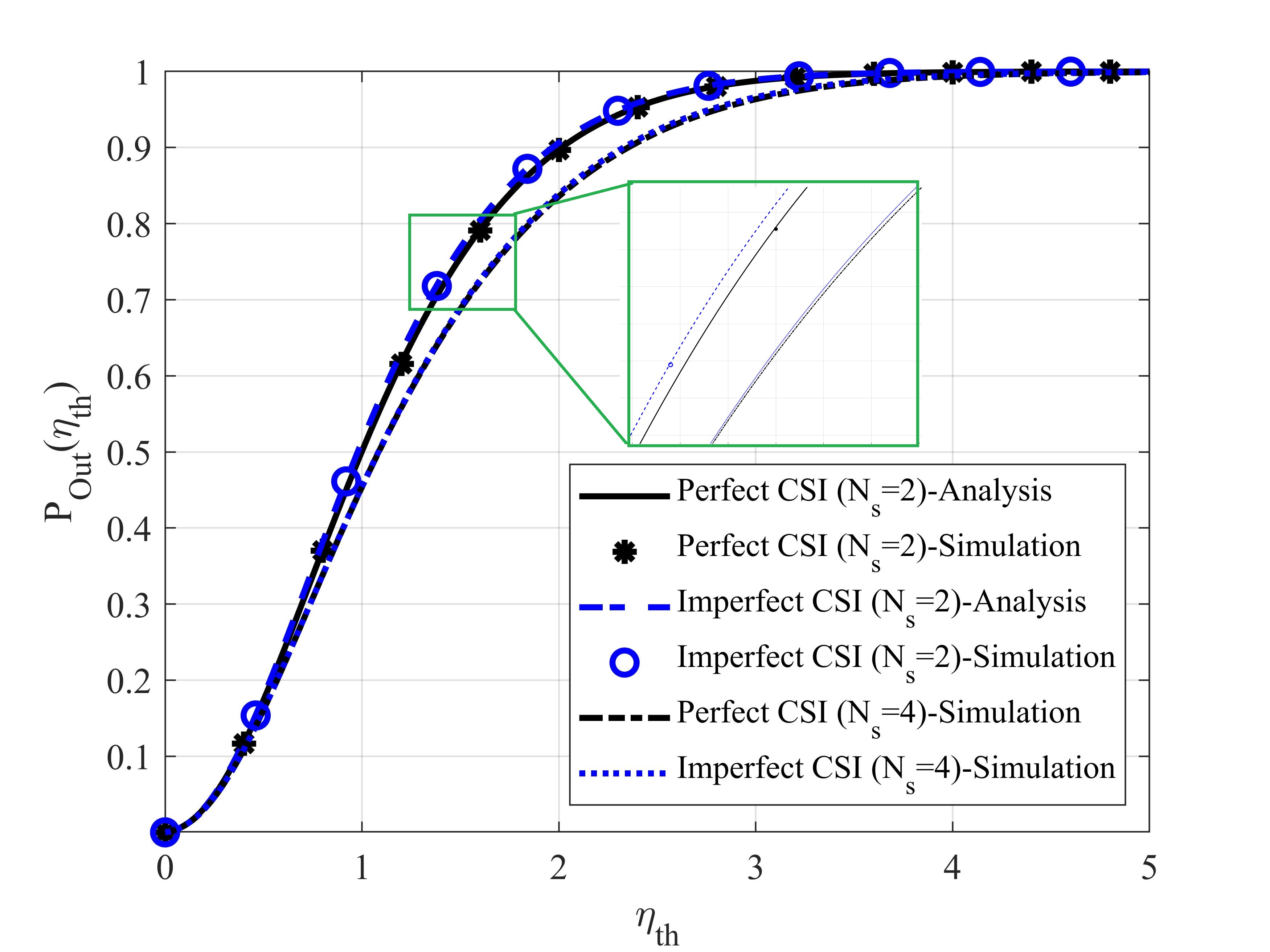}
		\caption{The outage probability versus threshold at the destination node when CSI is perfect or imperfect whit 2 or 4 active antenna at the source.}
		\label{fig3}
	\end{figure}

	In Figure \ref{fig3}, we plot the outage probability versus threshold at the destination node with perfect and imperfect CSI and 2 or 4 active antennae at the source. 
 The graph of perfect or imperfect CSI are very close to each other, but still perfect CSI has a lower outage probability for a given threshold. Also, the more active antennae the less outage probability.
	
\begin{figure}
		\centering
		\includegraphics[width=1.05\linewidth]{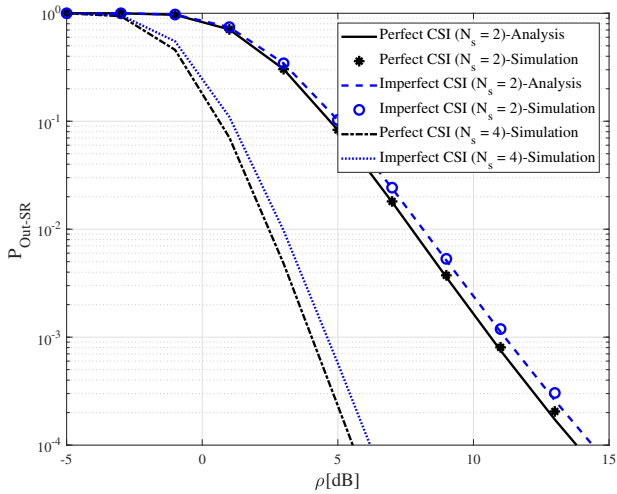}
		\caption{The outage probability versus average SNR at the relay node when CSI is perfect or imperfect whit 2 or 4 active antenna at the source.}
		\label{fig4}
	\end{figure}

\begin{figure}[t!]
		\centering
		\includegraphics[width=1.05\linewidth]{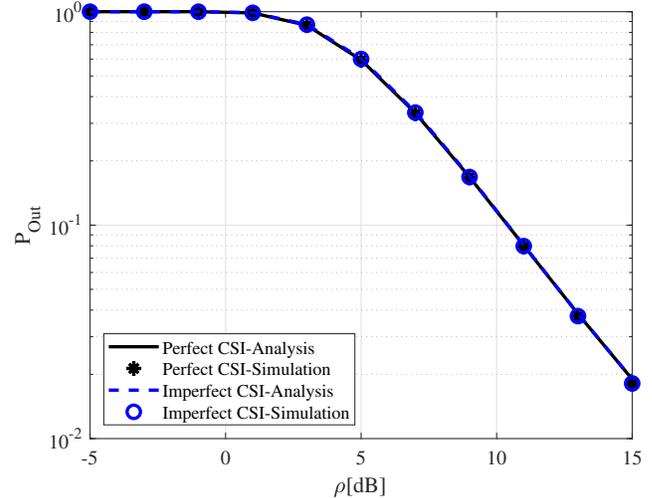}
		\caption{The outage probability versus average SNR at the destination node when CSI is perfect or imperfect whit 2 or 4 active antenna at the source.}
		\label{fig5}
	\end{figure}

	Finally, Figures \ref{fig4} and \ref{fig5} plot the outage probability versus the average SNR at the relay and destination node, respectively, with perfect and imperfect CSI. We can see the diversity orders of the outage probability are proportional to the number of antennae. 
 One interesting observation from Figure 5 is that the performance of the system in both modes, perfect and imperfect CSI, is very close to each other. This is useful in the case that the satellite wants to serve a large number of users and each user is assigned two antennae. To reduce the complexity, the estimation of the channel amplitude would be avoided in a practical implementation. The analytical results of this paper showed that the obtained formula has a simpler form when we have complete information, constituting a close approximation of the performance of non-ideal practical scenarios.
	
	\section{Conclusion}\label{s6}
	In this article, 
 a ground relaying scenario for satellite systems has been analyzed. 
 The obstructed connection between the source satellite and the ground destination is supported by an inactive ground user acting as a relay. Exact closed-form relationships of the SNR and the outage probability were obtained for the case where the satellite has two transmitting antennas. The analysis is supported by numerical evaluations. If the perfect CSI is not available in the pre-coding and only the signal phase is estimated, the performance loss is not too high and the system can reach its ideal performance by spending  limited power.

	\bibliographystyle{IEEEtran}
	\bibliography{ReferencesBIB}

\begin{thebibliography}{10}
\providecommand{\url}[1]{#1}
\csname url@samestyle\endcsname
\providecommand{\newblock}{\relax}
\providecommand{\bibinfo}[2]{#2}
\providecommand{\BIBentrySTDinterwordspacing}{\spaceskip=0pt\relax}
\providecommand{\BIBentryALTinterwordstretchfactor}{4}
\providecommand{\BIBentryALTinterwordspacing}{\spaceskip=\fontdimen2\font plus
\BIBentryALTinterwordstretchfactor\fontdimen3\font minus
  \fontdimen4\font\relax}
\providecommand{\BIBforeignlanguage}[2]{{%
\expandafter\ifx\csname l@#1\endcsname\relax
\typeout{** WARNING: IEEEtran.bst: No hyphenation pattern has been}%
\typeout{** loaded for the language `#1'. Using the pattern for}%
\typeout{** the default language instead.}%
\else
\language=\csname l@#1\endcsname
\fi
#2}}
\providecommand{\BIBdecl}{\relax}
\BIBdecl

\bibitem{3GPPTR38.811}
3GPP, ``{Study on New Radio ({NR}) to support non-terrestrial networks },''
  \emph{TR 38.811 V15.1.0}, Jun. 2019.

\bibitem{3GPPTR22.822}
3GPP, ``{Study on using Satellite Access in {5G}},'' \emph{TR 22.822 V16.0.0},
  Jun. 2018.

\bibitem{TR38.913}
3GPP, ``Study on scenarios and requirements for next generation access
  technologies,'' \emph{TR 38.913 V16.0.0}, May 2020.

\bibitem{3GPPTR38.821}
3GPP, ``{Solutions for {NR} to support non-terrestrial networks {(NTN)}},''
  \emph{TR 38.821 V16.0.0}, Dec. 2019.

\bibitem{9328305}
B.~Ji, Y.~Wang, K.~Song, C.~Li, H.~Wen, V.~G. Menon, and S.~Mumtaz, ``{A Survey
  of Computational Intelligence for 6G: Key Technologies, Applications and
  Trends},'' \emph{IEEE Transactions on Industrial Informatics}, vol.~17,
  no.~10, pp. 7145--7154, 2021.

\bibitem{7109864}
T.~S. Rappaport, G.~R. MacCartney, M.~K. Samimi, and S.~Sun, ``{Wideband
  Millimeter-Wave Propagation Measurements and Channel Models for Future
  Wireless Communication System Design},'' \emph{IEEE Transactions on
  Communications}, vol.~63, no.~9, pp. 3029--3056, 2015.

\bibitem{laneman2004}
J.~Laneman, D.~Tse, and G.~Wornell, ``Cooperative diversity in wireless
  networks: Efficient protocols and outage behavior,'' \emph{IEEE Transactions
  on Information Theory}, vol.~50, no.~12, pp. 3062--3080, 2004.

\bibitem{conne2010}
C.~Conne and I.-M. Kim, ``Outage probability of multi-hop amplify-and-forward
  relay systems,'' \emph{IEEE Transactions on Wireless Communications}, vol.~9,
  no.~3, pp. 1139--1149, 2010.

\bibitem{9789274}
H.~Zhang, G.~Pan, S.~Ke, S.~Wang, and J.~An, ``{Outage Analysis of Cooperative
  Satellite-Aerial-Terrestrial Networks With Spatially Random Terminals},''
  \emph{IEEE Transactions on Communications}, vol.~70, no.~7, pp. 4972--4987,
  2022.

\bibitem{9789278}
H.~Zhang, C.~Du, S.~Wang, G.~Pan, and J.~An, ``{Effects of Spatially Random
  Space Interference on Satellite-Aerial Downlink Transmission},'' \emph{IEEE
  Transactions on Communications}, vol.~70, no.~7, pp. 4956--4971, 2022.

\bibitem{9776485}
T.~N. Nguyen, L.-T. Tu, D.-H. Tran, V.-D. Phan, M.~Voznak, S.~Chatzinotas, and
  Z.~Ding, ``{Outage Performance of Satellite Terrestrial Full-Duplex Relaying
  Networks With co-Channel Interference},'' \emph{IEEE Wireless Communications
  Letters}, vol.~11, no.~7, pp. 1478--1482, 2022.

\bibitem{9774023}
W.~Cao, Y.~Zou, Z.~Yang, B.~Li, Y.~Lin, Y.~Li, W.~Wu, and L.~Liu, ``{Secrecy
  Outage Analysis of Relay-User Pairing for Secure Hybrid Satellite-Terrestrial
  Networks},'' \emph{IEEE Transactions on Vehicular Technology}, vol.~71,
  no.~8, pp. 8906--8918, 2022.

\bibitem{9744456}
L.~Han, W.-P. Zhu, and M.~Lin, ``{Outage Analysis of Multi-Relay NOMA-Based
  Hybrid Satellite-Terrestrial Relay Networks},'' \emph{IEEE Transactions on
  Vehicular Technology}, vol.~71, no.~6, pp. 6469--6487, 2022.

\bibitem{8747458}
X.~Zhang, B.~Zhang, K.~An, Z.~Chen, S.~Xie, H.~Wang, L.~Wang, and D.~Guo,
  ``{Outage Performance of NOMA-Based Cognitive Hybrid Satellite-Terrestrial
  Overlay Networks by Amplify-and-Forward Protocols},'' \emph{IEEE Access},
  vol.~7, pp. 85\,372--85\,381, 2019.

\bibitem{7917287}
J.~M. Romero-Jerez, F.~J. Lopez-Martinez, J.~F. Paris, and A.~J. Goldsmith,
  ``{The Fluctuating Two-Ray Fading Model: Statistical Characterization and
  Performance Analysis},'' \emph{IEEE Transactions on Wireless Communications},
  vol.~16, no.~7, pp. 4420--4432, 2017.

\bibitem{1198102}
A.~Abdi, W.~Lau, M.-S. Alouini, and M.~Kaveh, ``A new simple model for land
  mobile satellite channels: first- and second-order statistics,'' \emph{IEEE
  Transactions on Wireless Communications}, vol.~2, no.~3, pp. 519--528, 2003.

\bibitem{7503970}
M.~K. Samimi, G.~R. MacCartney, S.~Sun, and T.~S. Rappaport, ``{28 GHz
  Millimeter-Wave Ultrawideband Small-Scale Fading Models in Wireless
  Channels},'' in \emph{2016 IEEE 83rd Vehicular Technology Conference (VTC
  Spring)}, 2016, pp. 1--6.

\bibitem{HASHEMI2020100991}
\BIBentryALTinterwordspacing
H.~Hashemi, J.~Haghighat, and M.~Eslami, ``Performance analysis of relay-aided
  millimeter-wave communications with optimal and suboptimal combining at
  destination,'' \emph{Physical Communication}, vol.~39, p. 100991, 2020.
  [Online]. Available:
  \url{https://www.sciencedirect.com/science/article/pii/S1874490719304471}
\BIBentrySTDinterwordspacing

\bibitem{9201014}
H.~Hashemi, J.~Haghighat, M.~Eslami, and K.~Navaie, ``Amplify-and-forward
  relaying with maximal ratio combining over fluctuating two-ray channel:
  Non-asymptotic and asymptotic performance analysis,'' \emph{IEEE Transactions
  on Communications}, vol.~68, no.~12, pp. 7446--7459, 2020.

\bibitem{zwillinger2007table}
\BIBentryALTinterwordspacing
D.~Zwillinger and A.~Jeffrey, \emph{Table of Integrals, Series, and
  Products}.\hskip 1em plus 0.5em minus 0.4em\relax Elsevier Science, 2007.
  [Online]. Available: \url{https://books.google.com/books?id=aBgFYxKHUjsC}
\BIBentrySTDinterwordspacing

\end{thebibliography}

\end{document}